\documentclass[aps, prd, reprint, superscriptaddress]{revtex4-2}

\usepackage{amsmath}
\usepackage{graphicx}
\usepackage{subfigure}
\usepackage{hyperref}
\usepackage{float}

\begin{document}


\title{Characterisation at the HL-LHC of Long-lived Heavy Neutrinos \\ in Gauge Extensions of the Standard Model}


\author{Aram Fatah}
\email[]{aram.fatah.6597@student.uu.se}
\affiliation{Department of Physics \& Astronomy, Uppsala University, Box 516, 75120 Uppsala, Sweden}

\author{Stefano Moretti}
\email[]{stefano.moretti@physics.uu.se}
\email[]{s.moretti@soton.ac.uk}
\affiliation{Department of Physics \& Astronomy, Uppsala University, Box 516, 75120 Uppsala, Sweden}
\affiliation{School of Physics \& Astronomy, University of Southampton, Highfield, Southampton SO17 1BJ, United Kingdom}

\author{Antons Nikolajevs}
\email[]{antons.nikolajevs.1911@student.uu.se}
\email{antons.nikolajevs@uni-bielefeld.de}
\affiliation{Department of Physics \& Astronomy, Uppsala University, Box 516, 75120 Uppsala, Sweden}
\affiliation{Fakult\"at f\"ur Physik, Universit\"at Bielefeld, 33501 Bielefeld, Germany}


\begin{abstract}
We show how signals of heavy neutrinos with displaced decays can be detected at the Large Hadron Collider in two theoretical setups, both exploiting extended gauge sectors as portals to such new physics, 
the Left-Right Symmetric Model and $U(1)_{B-L}$. Further, owing to the reduced contamination from backgrounds away from the interaction point, we illustrate how the properties of the heavy neutrinos (mass, width and quantum numbers) can neatly be accessed at the High-Luminosity upgrade of the CERN machine. 
\end{abstract}


\maketitle

\section{Introduction}

We know of several effects of Beyond the Standard Model (BSM) physics, embodying feeble interactions, that can lead to extremely small decay widths, which in turn means long-lived particles. Here we shall build up cases where models of neutrino mass generation could lead to observable signals of Long-Lived Particles (LLPs) 
 at the forthcoming High-Luminosity Large Hadron Collider (HL-LHC) \cite{Gianotti:2002xx} or some foreseeable future collider.

Neutrino oscillation experiments \cite{Super-Kamiokande:1998kpq,LSND:2001aii,K2K:2002icj,T2K:2011ypd,DayaBay:2013yxg} have provided strong evidence in favour of at least two neutrino flavours being massive. As the neutrinos are so much lighter than any other SM fermions, the small masses are thought to be the result of some kind of seesaw mechanism. Perhaps the simplest realisation of such an idea is to add right-handed neutrinos and a Majorana mass term for these. The problem with the ensuing simple type-I seesaw model is that the production rate for heavy neutrinos is proportional to the square of the mixing between the left- and right-handed neutrinos, $|V_{\ell N}|^{2}$, which is too small to result in any events in the mass range where right-handed neutrinos are expected to decay within a typical HL-LHC detector.

Therefore, if we wish to study heavy neutrinos at such a collider, which is our phenomenological context here, we have to hope that Nature has chosen a non-minimal way of producing neutrino masses and thus consider models beyond the simplest seesaw extensions. Gauge extensions of the SM can provide new portals for the production of the heavy neutrinos. In fact, many of the early proposals for a seesaw mechanism \cite{Minkowski:1977sc,Mohapatra:1979ia,Mohapatra:1980yp,Mohapatra:1980qe,Wetterich:1981bx} were formulated in a left-right symmetric framework, where the smallness of neutrino masses is related to the spontaneous breaking of parity and lepton number. Gell-Mann, Ramond, Slansky \cite{Gell-Mann:1979vob} and others \cite{Magg:1980ut,Vergados:1985pq} connected the smallness of the neutrino masses more generally to a hierarchy of scales in the spontaneous breaking of a larger gauge group in Grand Unified Theories (GUTs).

In this work we consider two models, where right-handed neutrinos are necessary, the Left-Right Symmetric Model (LRSM) \cite{Pati:1974yy,Mohapatra:1974hk,Senjanovic:1975rk} and the $U(1)_{B-L}$ model \cite{Buchmuller:1991ce,Khalil:2006yi}. In the former, right-handed neutrinos are needed to ensure the parity invariance of the Lagrangian while, in the latter, they are necessary to avoid gauge anomalies. As we shall discuss below, in both cases, one can achieve a reasonable event rate at TeV-scale colliders for right-handed neutrino production if the symmetry breaking scale is not too large in comparison to the
relevant dynamical scale.

In LRSMs, long-lived neutrino production has been studied both at hadron colliders \cite{Cottin:2018kmq,Nemevsek:2018bbt} and future lepton colliders \cite{Urquia-Calderon:2023dkf}. For the $U(1)_{B-L}$ model there exist studies of long-lived neutrinos at hadron colliders \cite{Basso:2008iv,Bandyopadhyay:2022mej}. Both ATLAS \cite{ATLAS:2025uah} and CMS \cite{CMS:2024hik} have searched for displaced heavy neutrinos with  results so far.

Herein, under the assumption of a BSM gauge portal for heavy neutrino production, we aim at surpassing previous results from literature \cite{Cottin:2018kmq,Nemevsek:2018bbt,Urquia-Calderon:2023dkf,Basso:2008iv,Bandyopadhyay:2022mej} in three directions. Firstly, we simultaneously study neutral and charged current production of 
heavy neutrinos by adopting masses for the new gauge bosons compatible with the latest experimental constraints (in particular, using a redefinition of the relevant signals that include interference effects with the SM). Secondly, we individuate the non-excluded regions of parameter space in both the LRSM and $U(1)_{B-L}$ model where heavy neutrinos are LLPs producing displaced vertices inside the ATLAS and CMS detectors in phase space regions that are essentially free from SM backgrounds. Thirdly, we show how one can use both kinematic properties and quantum numbers of the ensuing leptons and jets to define observables (differential cross sections and their asymmetries) that can be used for characterising both the portal gauge bosons and  heavy neutrinos.  
However, as our purpose here is to provide an initial proof-of-concept of the feasibility of our studies, we limit ourselves to a partonic analysis.

The plan of this paper is as follows. In the next section, we discuss the aforementioned BSM scenarios. Then, we describe the analysis that we will deploy to extract signals of LLPs in the form of heavy neutrinos at the HL-LHC. After this, we present our results and conclude.

\section{The Models}

\subsection{LRSM}

The fundamental idea of LRSMs is that parity is a symmetry of Nature, which is then spontaneously broken. To incorporate this idea, the SM gauge group needs to be extended to $SU(3)_C\times SU(2)_L\times SU(2)_R\times U(1)_{B-L}$ \cite{Pati:1974yy,Mohapatra:1974hk,Senjanovic:1975rk}.  This group must then be broken to the SM gauge group through an extended scalar sector. We choose triplets $\Delta_{L,R}$ of SU$(2)_{L,R}$ to break the left-right symmetry. The Electro-Weak (EW) symmetry will then be broken through a bi-doublet field $\phi$. Each multiplet has the following representation and vacuum state:

\begin{eqnarray}
    \Delta_L &&= \begin{pmatrix}
            \delta_L^+/\sqrt{2} & \delta_L^{++} \\ \delta_L^0 & -\delta_L^+/\sqrt{2} 
        \end{pmatrix}  \nonumber \\ &&\to  \begin{pmatrix}
            0 & 0 \\ u_L & 0
        \end{pmatrix}\sim (3, 1, 2),\\
    \Delta_R &&= \begin{pmatrix}
            \delta_R^+/\sqrt{2} & \delta_R^{++} \\ \delta_R^0 & -\delta_R^+/\sqrt{2} 
        \end{pmatrix}  \nonumber \\ &&\to \begin{pmatrix}
            0 & 0 \\ u_R & 0
        \end{pmatrix} \sim (1, 3, 2),\\
    \phi &&= \begin{pmatrix}
            \phi_1^0 & \phi_2^+\\\phi_1^- & \phi_2^0
        \end{pmatrix} \to \begin{pmatrix}
            v & 0 \\ 0 & w
        \end{pmatrix} \sim (2, 2, 0).
\end{eqnarray}

The neutral components of the multiplets acquire Vacuum Expectation Values (VEVs), as a result the gauge group is spontaneously broken by $\Delta_R$ to the SM gauge group and then further broken by the bi-doublet $\phi$ and the triplet $\Delta_L$ to $U(1)_{\text{EM}}$. As a result of such a symmetry breaking, some of the gauge bosons become massive. In the limit of $u_R \gg v,w\gg u_L$, the masses are:
\begin{eqnarray}
    W &&= W_L \cos\zeta + W_R \sin\zeta,\\
    M_{W}^2 &&\approx \frac{1}{4}g^2 (v^2 + w^2),\\
    W' &&= -W_L \sin\zeta + W_R \cos\zeta,\\
    M_{W'}^2 &&\approx \frac{1}{4}g^2 (2u_R^2 + v^2 + w^2),
\end{eqnarray}
where $W$ and $W'$ are the mass eigenstates\footnote{As the mixing between the $W$ and $W'$ is tiny, we neglect considering it throughout.}, which are linear combinations of left- and right-handed charged $W_{L/R}$ fields. The lightest mass eigenstate $W$ can be identified as the mostly left-handed SM-like $W$ boson while the $W'$ is the heavy, mostly right-handed charged boson. The mixing angle $\zeta$ is defined as $\tan 2\zeta = \frac{2vw}{u_R^2 - u_L^2}$. Since weak interactions are of $V-A$ type, to the extent that they can experimentally be probed, either one of the VEVs, $v$ or $w$, needs to be small or $u_{R}$ must be very large.

Similarly, the neutral bosons have the following masses and mixings:
\begin{eqnarray}
        A =&& \sin\theta_W (A^3_L + A^3_R) + \sqrt{\cos2\theta_W}B,\\
        M_A^2 =&& 0,\\
        Z =&& \cos\theta_W A^3_L - \sin\theta_W \tan\theta_W A_R^3 \nonumber\\
        &&- \tan \theta_W \sqrt{\cos 2\theta_W}B\\
        M_Z^2 \approx&& \frac{g^2}{2\cos^2\theta_W} (v^2 + w^2 + 4u_L^2),\\
        Z' =&& \frac{\sqrt{\cos2\theta_W}}{\cos\theta_W} A^3_R - \tan\theta_W B\\
        M_{Z'}^2 \approx&& \frac{2g^2 \cos^2\theta_W}{\cos2\theta_W} u_R^2,
\end{eqnarray}
where $A$ is a massless photon, $Z$ is the SM-like $Z$ boson and $Z'$ is the new heavy neutral boson. We always have $m_{Z^{\prime}}>m_{W^{\prime}}$, so the $W^{\prime}$ state will be the main portal for our long-lived neutrino production. The $W^{\prime}$ has been searched for in the dijet \cite{CMS:2019gwf} and lepton and heavy neutrino \cite{CMS:2021dzb,ATLAS:2023cjo} modes. The dijet bound is independent of the  model details and leads to $m_{W^{\prime}}>3.6$~TeV, while the leptonic mode has a stronger bound, but it depends on the right-handed neutrino mass and assumes a prompt neutrino decay.

Contrary to the SM, both left- and right-handed fermions are assembled into $SU(2)$ doublets:
\begin{equation}
    L_{L/R}^i = \begin{pmatrix}
        \nu_i \\ \ell_i
    \end{pmatrix}_{L/R},\qquad Q^i_{L/R} = \begin{pmatrix}
        u_i \\ d_i
    \end{pmatrix}_{L/R},
\end{equation}
where $i$ is the flavour index and $\nu_R$ is the right-handed neutrino
(which is absent in the SM). Yukawa interactions of fermionic doublets with the scalar multiplets give rise to fermion masses. Interaction with the bi-doublet allows for a Dirac mass term, for quarks and leptons, while interaction with the triplets produces a Majorana mass term for the left- and right-handed neutrinos, respectively. The corresponding Lagrangians are:
\begin{eqnarray}
    \mathcal{L}_{\text{Dirac}} =&& Y_{ij}^\ell\bar{L}_L^i \phi L^i_R + \Gamma_{ij}^\ell \bar{L}^i_L \tilde{\phi} L_R^j\nonumber\\
    &&+Y^q_{ij}Q_L^i \phi Q^j_R+\Gamma_{ij}^q Q_L^i \tilde{\phi} Q_R^j + \operatorname{h.c.},
\end{eqnarray}
\begin{eqnarray}
    \mathcal{L}_{\text{Majorana}} = &&-h_{ij}^L\,\text{i}\,\bar{L}_L^{c i}\,\sigma_2\, \Delta_L\,L_L^j\nonumber\\
    &&-h_{ij}^R\,\text{i}\,\bar{L}_R^{ci}\,\sigma_2 \Delta_R\,L_R^j \;.
\end{eqnarray}
After  symmetry breaking, quarks and charged leptons acquire the following masses:
\begin{eqnarray}
    M_u &&= \frac{1}{\sqrt{2}} (Y^q_{ij} v + w \Gamma^q_{ij})\\
    M_d &&= \frac{1}{\sqrt{2}} (Y^q_{ij} w + v \Gamma^q_{ij})\\
    M_\ell &&= \frac{1}{\sqrt{2}} (Y^\ell_{ij} w + v \Gamma^\ell_{ij})\;.
\end{eqnarray}
In contrast, neutrino mass generation happens via a seesaw mechanism. The large right-handed scalar triplet VEV, $u_R$,  creates a Majorana mass for the right-handed neutrino via type-II seesaw. Thereafter left-handed neutrinos get a mass via type-I seesaw. In principle one can generate a mass for left-handed neutrinos through the left-handed triplet VEV and the type-II seesaw. However, the left-handed triplet VEV is heavily constrained, as a result it is often set to zero.  The resulting mass matrix is given by
\begin{equation}
    \begin{pmatrix}
        0 & M_D \\ M_D^T & M_R
    \end{pmatrix},
\end{equation}
where $M_D$ is the Dirac mass, given by $M_D =\frac{1}{\sqrt{2}} (Y^\ell w + \Gamma^\ell v)$, and $M_R$ is the Majorana mass matrix, given by  $M_R = \sqrt{2}h_R u_R$. After diagonalisation, the neutrino mass states are $\nu$, and $N$, with the following masses:
\begin{equation}
    M_N \approx M_R,\qquad M_\nu = -M_D M_R^{-1} M_D^\dagger.
\end{equation}
The mostly left-handed SM-like neutrino $\nu$ stays light if the heavy neutrino $N$ has a sufficiently large mass.

\subsection{$U(1)_{B-L}$}
The second model that was considered is the $U(1)_{B-L}$ gauge extension of the SM, in which the difference between the baryon and lepton numbers is promoted to a local symmetry. The gauge group of the model is given by $SU(3)_C\times SU(2)_L \times U(1)_{Y} \times U'(1)_{B-L}$ \cite{Khalil:2006yi, Basso:2008iv}. The gauge anomaly is cancelled by introducing thee generations of right-handed neutrinos. Two Abelian groups,  $U(1)_{Y}$, and $U'(1)_{B-L}$, allow for the kinetic mixing term in the Lagrangian:
\begin{equation}
    \mathcal{L}_{\text{kin.}} \supset -\frac{1}{4}B_{\mu \nu} B^{\mu \nu} -\frac{1}{4}X_{\mu \nu} X^{\mu \nu} + \eta B_{\mu \nu} x^{\mu \nu},
\end{equation}
where $B_{\mu \nu}$ and $X_{\mu \nu}$ are the field strength tensors for the $U(1)_Y$ and $U'(1)_{B-L}$ gauge bosons, correspondingly. The gauge fields can be rotated in order to remove the kinetic mixing term, which would result in non-diagonal couplings in the covariant derivative:
\begin{equation}
    D_\mu = \partial_\mu +\text{i}g\tau^a A^a_\mu + \text{i}g' YB_\mu + \text{i}(\tilde{g}_1 Y + \tilde{g} (B-L))Z'_\mu.
\end{equation}
As a result, the ensuing $Z'$ gauge boson couples not only to $B-L$ but also to hypercharge. However, the non-diagonal coupling $\tilde{g}_1$ is constrained by the EW precision tests to be $\tilde{g}_1 < 10^{-2}$ \cite{Liu:2020nqi}. The scalar sector of the model consists of a SM-like Higgs $SU(2)_L$ doublet $H$ as well as a complex $SU(2)_L$ singlet scalar $\chi$. After the new symmetry breaking, the neutral components of the doublet as well as the singlet acquire a VEV:
\begin{equation}
    \langle H \rangle = \begin{pmatrix}
        0 \\ \frac{v}{\sqrt{2}}
    \end{pmatrix} \sim (1, 2, 1/2, 0),\quad \langle X \rangle = \frac{x}{\sqrt{2}} \sim (1, 1, 0, 2).
\end{equation}
The scalar singlet generates the mass of the $Z'$ boson: i.e., 
\begin{equation}
M_{Z'} = 2\tilde{g}x. 
\end{equation}
In addition, it generates a Majorana mass term for the right-handed neutrinos:
\begin{eqnarray}
    \mathcal{L} \supset &&-Y^d_{ij} \bar{Q}^i_L H d^j_R - Y^u_{ij} \bar{Q}_L^i \tilde{H} u^i_R - Y^\ell_{ij} \bar{L}_L^i H e_R^j \nonumber \\
    &&- Y^\nu_{ij} \bar{L}_L^i \tilde{H} N_R^j - Y^M_{ij}\bar{N}_R^{ci} \chi N_R^j + \text{h.c.}
\end{eqnarray}
This results in a mass matrix for the left- and right-handed neutrinos $\nu$ and $N$:
\begin{equation}
    \begin{pmatrix}
        0 & M_D\\
        M_D^T & M_R
    \end{pmatrix},
\end{equation}
with $M_D = \frac{1}{\sqrt{2}}Y^\nu v$ and $M_R = \sqrt{2} Y^Mx$. After diagonalisation in the limit $M_D \ll M_R$, one obtains the heavy and light neutrino masses:
\begin{equation}
\begin{split}
    &m_N \sim \sqrt2 Y^M x,\\
    &m_\nu \sim  \frac{1}{2\sqrt{2}} Y^\nu (Y^M)^{-1} (Y^\nu)^T \frac{v^2}{x}.
\end{split}
\end{equation}

\begin{figure}
\centering
        
        
        
        
        
        
        
    \includegraphics[width=0.7\linewidth]{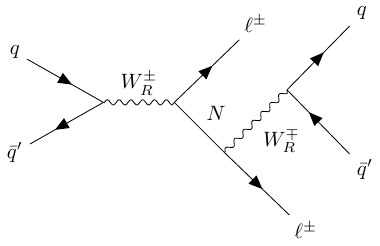}
    \caption{Feynman diagram for the process studied in the LRSM.}
    \label{fig:LRSM}
\end{figure}

\section{Analysis}
\subsection{Model Implementation}

The parameter space of each model was constrained using the computed  production cross section as measured by  ATLAS and CMS. The 
mass spectrum of each model was computed using \texttt{SPheno} \cite{Porod:2003um, Porod:2011nf}, a package that calculates particle masses, mixings and other observables based on the Lagrangian parameters. The 
model files for the $U(1)_{B-L}$ gauge extension were generated using \texttt{SARAH} \cite{Staub:2008uz, Staub:2009bi, Staub:2010jh, Staub:2012pb, Staub:2013tta}, a software that allows for supersymmetric and non-supersymmetric model building. For the LRSM, the model files provided  in Ref. \cite{Bonilla:2016fqd} were used. 
Finally, the production cross sections were calculated using \texttt{MadGraph5} \cite{Alwall:2014hca}. We carry out our analysis at parton level, using the latter's code default Parton Distribution Functions (PDFs) and dynamical renormalisation/factorisation scale. All our numerical computation are for $\sqrt s=13.6$ TeV.

\subsection{Collider Signatures}

\begin{figure}
    \centering

    \includegraphics[width=0.7\linewidth]{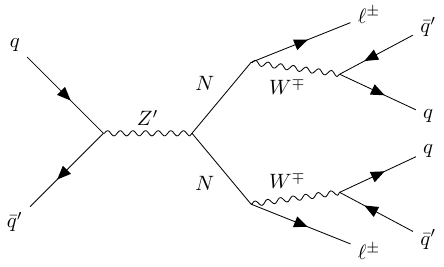}
    \caption{Feynman diagram for the process studied in $U(1)_{B-L}$.}
    \label{fig:U1BL}
\end{figure}

In the LRSM case, the process that we are targeting is
\begin{equation}
   q \bar q' \rightarrow W_R^\pm \rightarrow N_e l^{\pm} \rightarrow l^{\pm} l^\pm + \text{jets}
   \label{proc:LRSM}
\end{equation}
(see Fig.~\ref{fig:LRSM}) whereas in the $U(1)_{B-L}$ case we are studying 
\begin{equation}
q \bar q \rightarrow Z' \rightarrow N_e N_e \to l^{\pm} l^\pm + \text{jets}
\label{proc:U1BL}
\end{equation}
(see Fig.~\ref{fig:U1BL}), 
where $q^{(')}$ refers to a quark in the proton and $l=e,\mu$. (Hereafter, jets are intended as partons.)

\subsection{Signal Features}
In order to pursue studies of either BSM scenario introduced we need, on the one hand, to define the regions of parameter space not excluded by experiments (chiefly, ATLAS and CMS) and, on the other hand, define configurations of it yielding heavy neutrinos as LLPs. Therefore, we will first extract limits on the cross section for our target processes then introduce some Benchmark Points (BPs) amenable to phenomenological investigations. Our aim is to prove that the HL-LHC has sensitivity to these and we will do so by, firstly, computing the inclusive production and decay rates of the processes \ref{proc:LRSM}--\ref{proc:U1BL} and, secondly, studying some differential distributions enabling us the extract the properties (masses, widths/lifetimes and quantum numbers) of the $N_e$ states and, potentially, the $W'$ and $Z'$ 
ones.

\section{Results}
\subsection{LRSM}
The main constraint on the LRSM $W'$ and $N_e$ masses comes from the Keung-Senjanovi\'c (KS) process of Fig.~\ref{fig:LRSM} \cite{Keung:1983uu}. Searches have been performed by the ATLAS \cite{ATLAS:2015gtp, ATLAS:2018dcj, ATLAS:2019isd} and CMS \cite{CMS:2018agk, CMS:2021dzb} collaborations looking for $W'$ and $N_e$ precisely through the channel in Eq.~(\ref{proc:LRSM})\footnote{Notice that the SM contribution via a $s$-channel $W$ is here negligible, owing to the tiny mixing effects with the $W'$, so we have neglected this altogether. (For the same reason, we will refer to the new charged gauge boson of the LRSM as $W'$ and $W_R$ interchangeably.}. The results of these searches were used to constrain the parameter space of our LRSM model.

\begin{figure} [t]
    \centering
    \includegraphics[width=\linewidth]{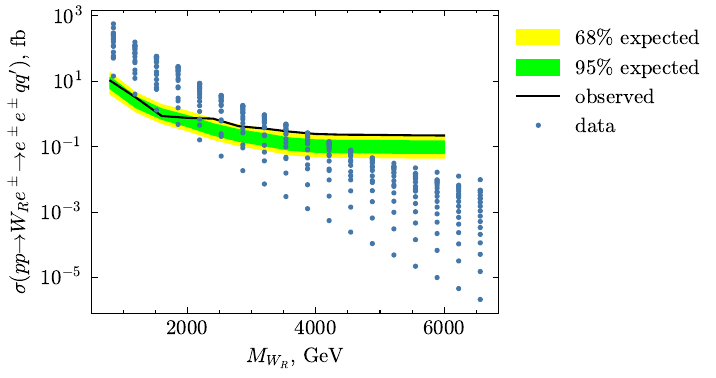}
    \caption{Comparison of the observed cross section (see Ref. \cite{CMS:2021dzb}) with the calculated one.}
    \label{fig:lrms-observed-cross section}
\end{figure}

\begin{figure} 
    \centering
    \subfigure[]{\includegraphics[width=0.45\textwidth]{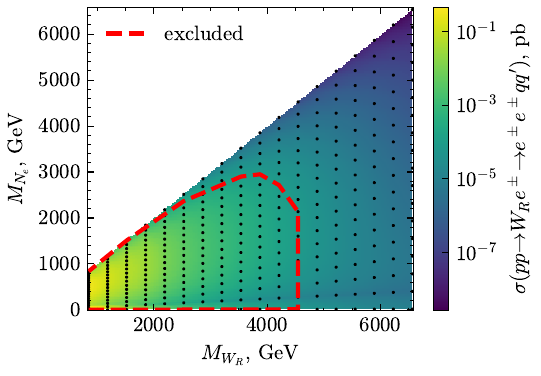}} 
    \subfigure[]{\includegraphics[width=0.45\textwidth]{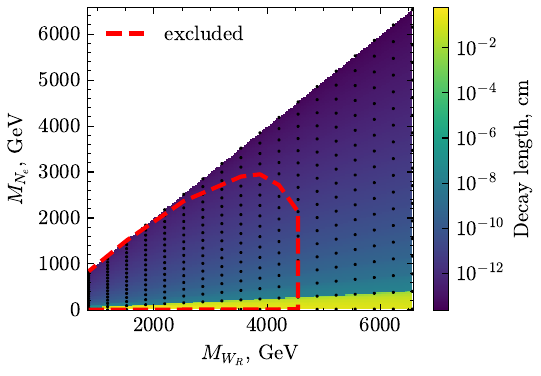}}
    \caption{(a) Calculated KS process cross section for different $W'$ and $N_e$ masses. (b) Heavy neutrino decay length for different $W'$ and $N_e$ masses. In both plots the excluded region is shown with a dashed line. }
    \label{fig:lrsm-param-space-scan-results}
\end{figure}

A scan over the Yukawa couplings, $h^R$, as well as the right-handed triplet VEV, $u_R$, was performed. In order to produce heavy neutrinos via the KS process, the requirement $M_N < M_{W'}$ has to be satisfied, which sets the upper limit on the Yukawa coupling $h^R < \frac{g}{2} \approx 0.33$, where $g$ is the $SU(2)_R$ gauge coupling. The scan was performed over the following values:
\begin{eqnarray}
    u_R && \in [1.8\times 10^{3},\;1.4\times10^{4}]~\mathrm{GeV}, \nonumber \\
    h^R_{11} && \in [1\times 10^{-3},\;0.33].
\end{eqnarray}
For each point a cross section for the $p p \to e^\pm e^\pm q q'$ channel was calculated and compared to the observed cross section given in Ref. \cite{CMS:2021dzb}, thus excluding the regions of parameter space  that are not consistent with such observations. Figs.~\ref{fig:lrms-observed-cross section} and \ref{fig:lrsm-param-space-scan-results} show the results of this exercise, highlighting the excluded parameter space region.


Three Benchmark Points (BPs) were selected from the surviving parameter space, where the heavy neutrino production cross sections are among the largest while also  having a wide range of heavy neutrino displacements. The main parameters of each BP are shown in Tab.~\ref{tab:lrsm-bps}. Current detectors are capable of reconstructing displaced vertices with a displacement as small as $\sim 100~\mathrm{\mu m}$ \cite{Ito:2017dpm}. The heavy neutrino $N_e$ is produced highly boosted due to being significantly lighter then the $W'$. As a result, the decay length of heavy neutrino in the laboratory frame is significantly larger as compared to the proper decay length. Heavy neutrinos, in BP1,  BP2 and BP3, have a decay length of $\sim 0.15$, $\sim 0.5~\mathrm{cm}$ and $\sim 0.04~\mathrm{cm}$, respectively, therefore they decay within the CMS tracker \cite{CMS:2013vyz} (hence the ATLAS one too).

\begin{table}
    \centering
    \begin{tabular}{cccc}\hline\hline
         & BP1 & BP2 & BP3\\\hline
         $M(Z')$, TeV& 7.93 & 7.93 & 7.93\\
         $M(W')$, TeV& 4.73 & 4.73 & 4.73\\
         $M(N_e)$, GeV& 22.1 & 17.1 & 28.6\\
         $\Gamma(N_e)$, GeV& $1.43\times 10^{-13}$ & $3.93\times 10^{-14}$ & $5.17\times 10^{-13}$\\
         $c\tau(N_e)$, cm & 0.138 & 0.504 & 0.038\\\hline\hline
    \end{tabular}
    \caption{Parameters of the BPs selected for the LRSM.}
    \label{tab:lrsm-bps}
\end{table}

\subsection{$U(1)_{B-L}$}
Similarly to the previous BSM scenario, the parameter space of the $U(1)_{B-L}$ gauge extension was then explored. The main constraints come from experimental searches in the dilepton ($e^+e^-$ and $\mu^+ \mu^-$) final state in the
Drell-Yan process \cite{ATLAS:2019erb} as well as in the diboson ($W^+ W^-$) final state \cite{ATLAS:2020fry}. The parameter space of this model was again  constrained by comparing the cross sections produced following a parameter scan\footnote{In general, when computing process (\ref{proc:U1BL}), one should include interference effects between the $Z'$, $Z$, and $\gamma$. In fact, when the $Z'$ width is large, interference effects could reduce the cross section, thus weakening the experimental limits \cite{Accomando:2013sfa}. However, in the parameter space region of interest, where the heavy neutrino is long-lived, the $Z'$ was found to be rather narrow, thus making the aforementioned interference effects very small. (Nonetheless, the latter were included in the analysis for completeness.)}. 

We proceeded as follows. 
Firstly, the $Z'$ mass as well as the $U(1)_{B-L}$ gauge couplings $\tilde{g}_1$ and $\tilde{g}$ were constrained. To this end, a parameter scan over the gauge couplings and the scalar singlet VEV $x$ was performed. The parameter ranges in the scan were:
\begin{eqnarray}
        \tilde{g} && \in [0.015, 0.8], \nonumber \\
        \tilde{g}_1 &&\in [3\times10^{-5}, 0.1],\\
        x &&\in [2250, 70000]~\mathrm{GeV}\nonumber.
\end{eqnarray}

In addition, a constraint on the $Z-Z'$ mixing angle $\theta'$ was imposed 
\cite{Accomando:2016sge, Accomando:2017qcs}: i.e., all points with $\theta' \approx \frac{\tilde{g}_1}{2}\frac{M_Z}{M_{Z'}^2 - M_{Z}^2} \gtrsim 10^{-3}$ were excluded.  

\begin{figure}[t]
    \centering
    \subfigure[]{\includegraphics[width=0.45\textwidth]{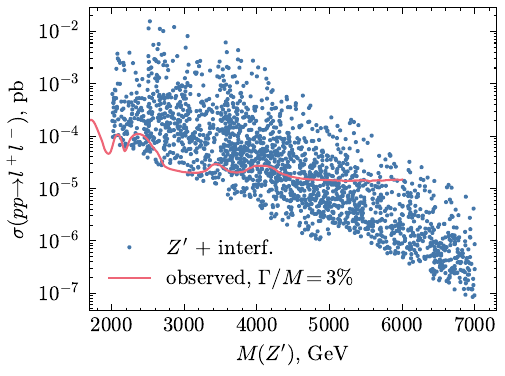}} 
    \subfigure[]{\includegraphics[width=0.45\textwidth]{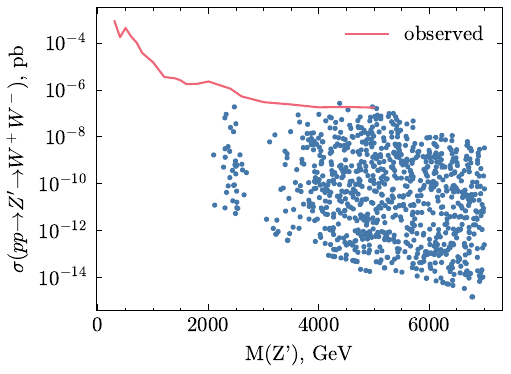}}
    \caption{(a) Calculated cross section for different parameter space points compared to the observed cross section in the dilepton final state. (b) Calculated cross section for different parameter space points compared to the observed cross section in the diboson final state (here, only the points that were not excluded by the dilepton limit are shown).}
    \label{fig:blsm-dilepton-diboson-limits}
\end{figure}

\begin{figure}[t]
    \centering
    \includegraphics[width=\linewidth]{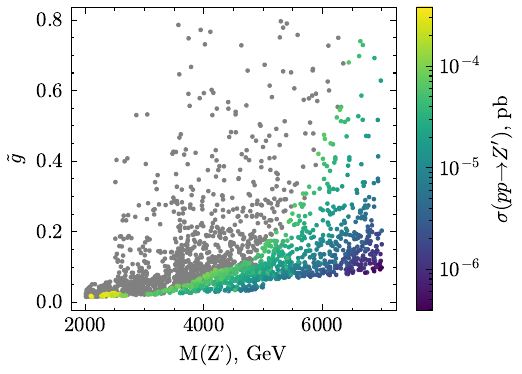}
    \caption{$Z'$ production cross sections for different $Z'$ masses and gauge couplings. Excluded points are shown in grey.}
    \label{fig:blsm-zp-scan}
\end{figure}

The results of the scan are shown in Fig.~\ref{fig:blsm-dilepton-diboson-limits}. Due to the small mixing between the $Z'$ and $Z$, the main contribution to the cross section of the $W^+ W^-$ production comes from the SM, thus excluding only a small part of the parameter space. The main constraint, instead, comes from the dilepton final state. Here, it was possible to evade the commonly reported mass limit of $\approx4.5~\mathrm{TeV}$ by lowering the gauge coupling $\tilde{g}$ while increasing the scalar singlet VEV. The production cross section of the $Z'$ boson for different masses and gauge couplings are summarised in Fig. \ref{fig:blsm-zp-scan}. 

\begin{figure}[!t]
    \centering
    \subfigure[]{\includegraphics[width=0.45\textwidth]{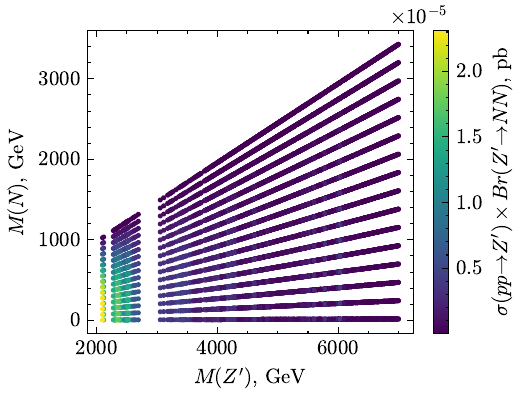}} 
    \subfigure[]{\includegraphics[width=0.45\textwidth]{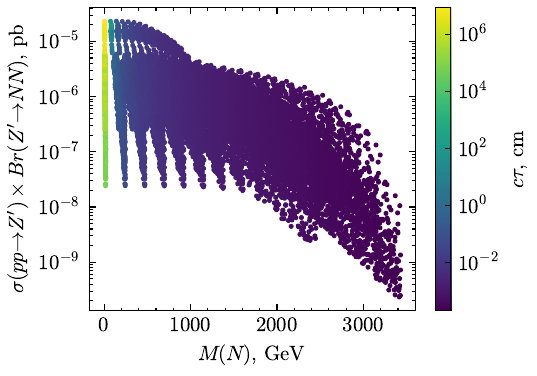}}
    \caption{(a) Heavy neutrino production cross sections for different $Z'$ and $N_e$ masses. (b) Heavy neutrino invariant lifetime for different $N_e$ masses.}
    \label{fig:blsm-N-scan}
\end{figure}

As we are interested in heavy neutrino production and displaced (from the interaction point) decays, we show in Fig. \ref{fig:blsm-N-scan} the cross sections for heavy neutrino pair production using the surviving parameter space points from Fig. \ref{fig:blsm-zp-scan}. We see that heavy neutrinos have a large production cross section and has a sizeable proper lifetime when the mass of the heavy neutrino is rather small.

Based on these results, again, three BPs with a wide range of decay lengths were selected for further analysis. Tab.~\ref{tab:blsm-bps} shows the main parameters for each BP. From here, one can notice that the BPs in $U(1)_{B-L}$
were chosen with displacements that are similar to those selected for LRSM. 

\begin{table}
    \centering
    \begin{tabular}{cccc}\hline\hline
         & BP1 & BP2 & BP3\\\hline
         $M(Z')$, TeV & 2.36  & 2.62  & 2.09 \\
         $M(N_e)$, GeV & 158.99  & 176.61 & 277.47 \\
         $\Gamma(N_e)$, GeV& $1.81\times 10^{-13}$ & $6.57\times 10^{-14}$ & $2.26\times 10^{-13}$\\
         $c\tau(N_e)$, cm & 0.150  & 0.109  & 0.036 \\\hline\hline
    \end{tabular}
    \caption{Parameters of the BPs selected for the $U(1)_{B-L}$ model.}
    \label{tab:blsm-bps}
\end{table}

\subsection{Phenomenological Results for the LRSM}
As intimated, a promising collision process to look at in colliders for the LRSM is the KS process depicted in Fig.~\ref{fig:LRSM}. Since, in our chosen BPs, one has $M({N_{e}}) < M_{W'}$, the KS process provides a clean signal that would confirm the Majorana nature of the heavy neutrinos as the final state leptons can be any combination of charges. Detecting such a process would show the lepton number violating behaviour of the heavy neutrinos. 
The choice of BPs for the process $p \space  p \xrightarrow{} W_R \xrightarrow{} N_e \space e^{\pm} \xrightarrow{} e^{\pm} \space e^{\pm} \space j \space j$ are  displayed in Tab.~\ref{crosssections}, together with the corresponding cross sections. 
\begin{table}[]
    \centering
    \begin{tabular}{cc}
    \hline \hline
            & Cross section, pb  \\
            \hline
        BP1 &  $6.231 \times 10^{-4}$  \\
        BP2 &  $6.273\times 10^{-4}$ \\
        BP3 &  $6.170\times 10^{-4}$  \\ 
        \hline \hline
    \end{tabular}
    \caption{The cross section for the KS process for each BP.}
    \label{crosssections}
\end{table}

An interesting feature for the chosen BPs and the mass of the $W'$ is the off-shell production of $W'$ contribution to the cross section. For $M_{W'} \approx 5$ TeV, the off-shell contribution will start to dominate the cross section and this effect is evident when analysing the final state objects. What one should expect from the KS process and given that $M_{W'} \gg M_{N_e}$ is one hard lepton reaching the detector while the other lepton originating from $N_e$ resides within a jet cone. However, since the off-shell production of $W'$ bosons are starting to dominate the cross section at the chosen mass for $W'$ there will be a significant fraction of events where two leptons are resolved by the detector in the final states. Therefore, for the analysis, we plot the results for two signal regions (SRs), SR1 and SR2, one where only one lepton is resolved and one where two leptons are resolved, respectively.
\begin{figure}
    \centering
    \subfigure[]
    {\includegraphics[width=0.45\textwidth]{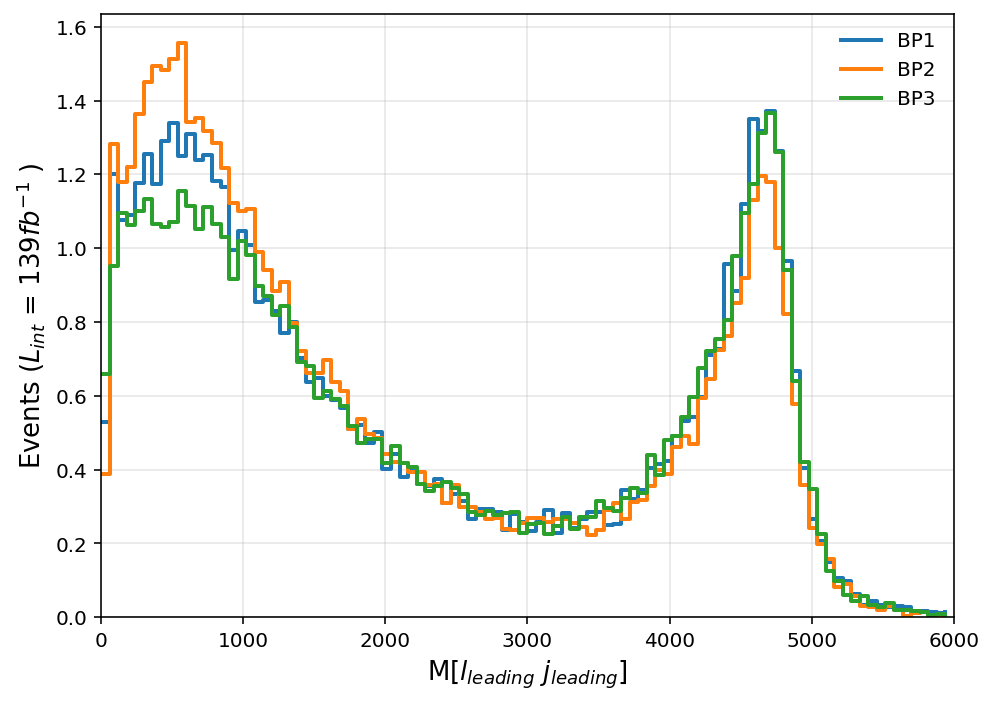}}
    \subfigure[]
    {\includegraphics[width=0.45\textwidth]{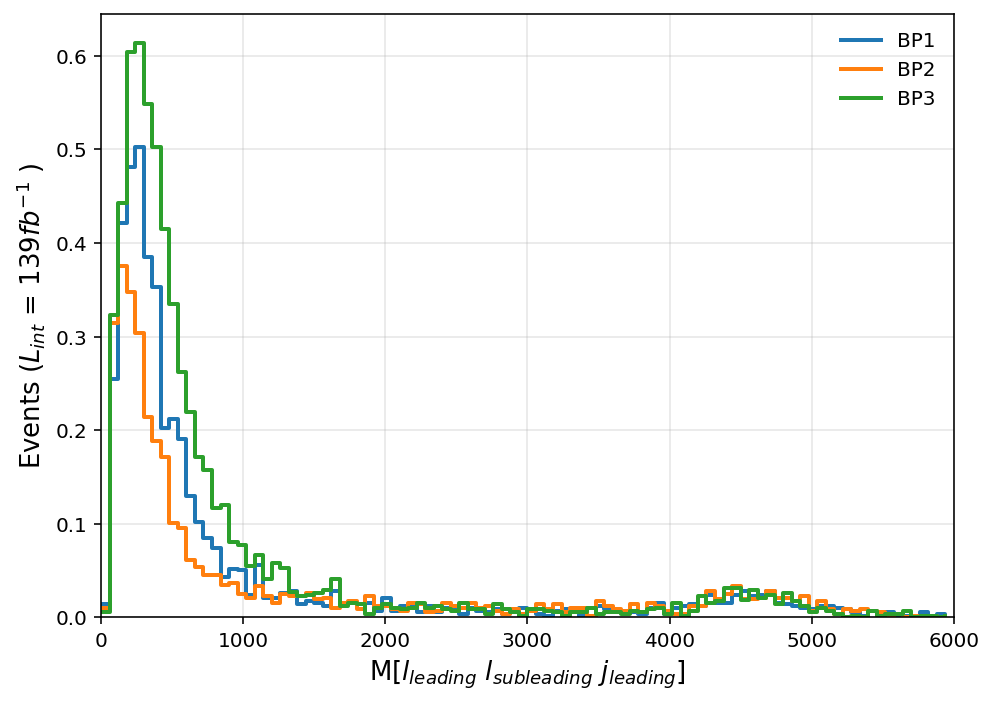}}
    \caption{(a) The invariant mass distribution (in GeV) for the leading lepton and leading jet is displayed for the three BPs in SR1. (b) The invariant mass distribution for the leading jet, leading and sub-leading leptons are displayed for the three BPs in SR2.  }
    \label{bp2_m_inv_w_sr1}
\end{figure}
The invariant mass distributions, normalised to an integrated luminosity of 139 fb$^{-1}$, for the three BPs are shown in Fig.~\ref{bp2_m_inv_w_sr1}. In particular, in Fig.~\ref{bp2_m_inv_w_sr1}(a), the invariant mass distribution for the leading lepton and jet is reconstructed for SR1 and in Fig.~\ref{bp2_m_inv_w_sr1}(b) the invariant mass distribution for the leading and sub-leading leptons and the leading jet is reconstructed for SR2.
The invariant mass distribution in Fig.~\ref{bp2_m_inv_w_sr1}(a) correctly reproduces the mass of $W'$ and at current LHC luminosities we should expect to produce only a few events where an on-shell $W'$ is produced for each of the BPs. In Ref.~\ref{bp2_m_inv_w_sr1}(b) we can see that the invariant mass distribution of the final state objects does not correctly reproduce the $W'$ mass as we would expect because, if two leptons are resolved in the final state, they most likely come from an off-shell $W'$. Therefore, we do not actually expect to see any events of $W'$ production where two final state leptons are resolved at current LHC luminosities. However, the more than tenfold increase in luminosity expected at the HL-LHC would render such events accessible therein.

\begin{figure}
    \centering
    \subfigure[]
    {\includegraphics[width=0.45\textwidth]{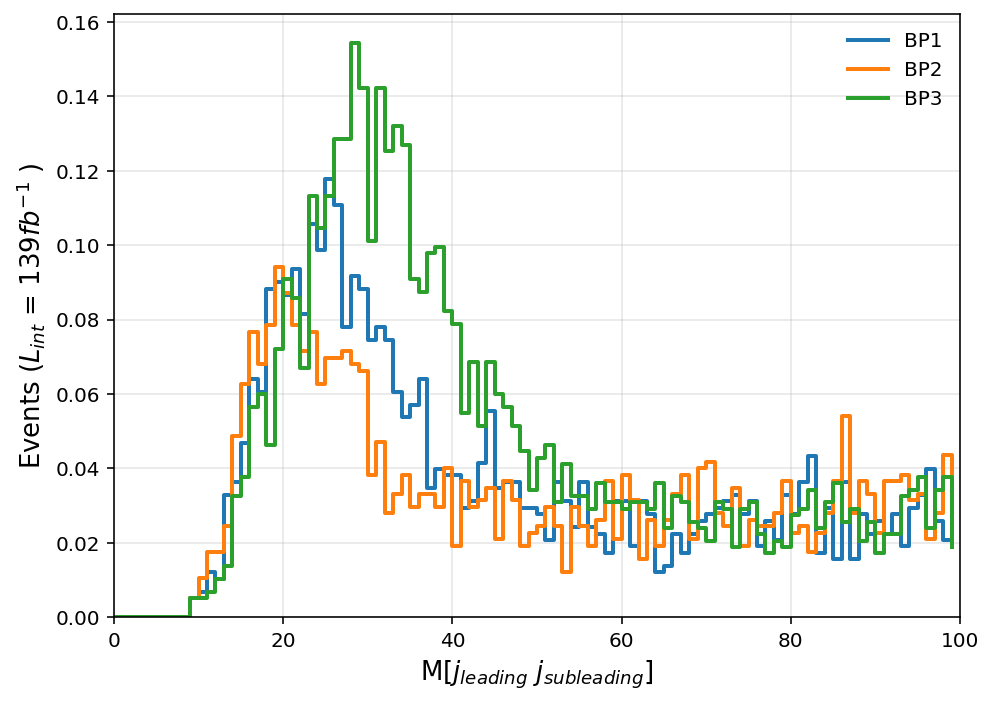}}
    \subfigure[]
    {\includegraphics[width=0.45\textwidth]{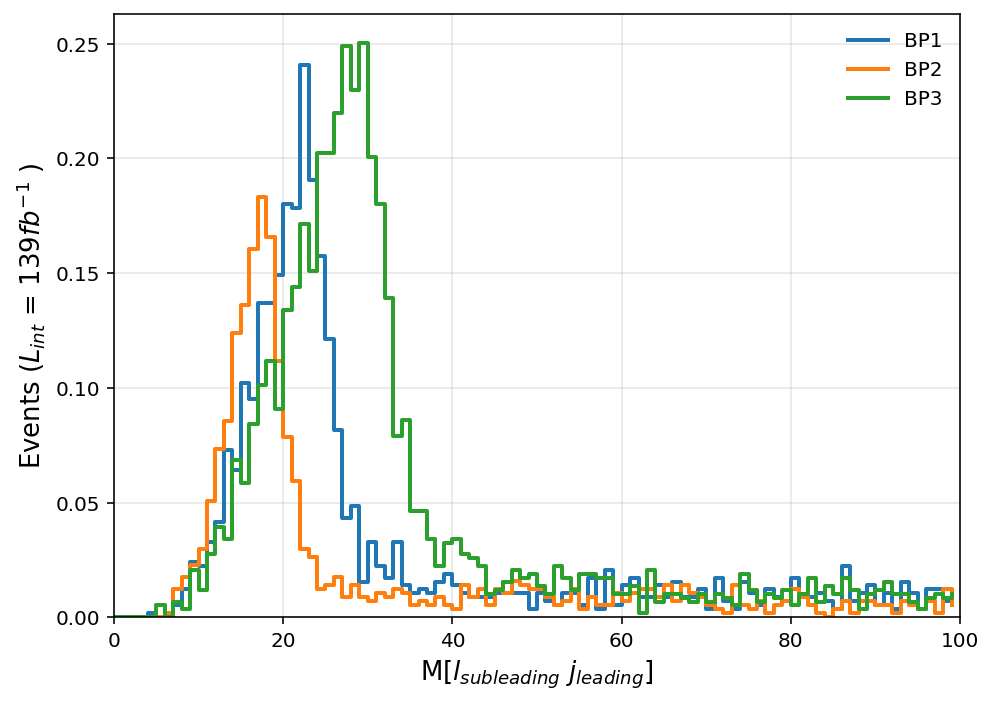}}
    \caption{(a) The invariant mass distribution (in GeV) for the leading jet and sub-leading jet is displayed for the three BPs in SR1. (b) The invariant mass distribution for the sub-leading lepton and sub-leading jet are displayed for the three BPs in SR2. }
    \label{bp2_m_inv_nu_sr1}
\end{figure}

To reconstruct the heavy neutrino mass we exclude the leading lepton from the invariant mass distribution plots, as it originates from the $W'$ boson. In Fig.~\ref{bp2_m_inv_nu_sr1}(a), we plot the leading and sub-leading jet distributions for SR1. The mean value of this distribution is around 600 GeV for each of the BPs, however, there is a clear peak around the respective heavy neutrino masses and we hardly expect to see around one event at current LHC luminosities. Similarly, in Fig.~\ref{bp2_m_inv_nu_sr1}(b), the invariant mass distribution for the sub-leading lepton and leading jet is shown. Since the majority of events in SR2 come from off-shell production of a $W'$, the final state objects will not be as boosted as in SR1 and therefore there is a much clearer peak around the respective heavy neutrino masses.  
Altogether, again, we do not expect to see any events at current LHC luminosities, yet scope for detection may exist at the HL-LHC.

\begin{figure}
    \centering
    \subfigure[]
    {\includegraphics[width=0.45\textwidth]{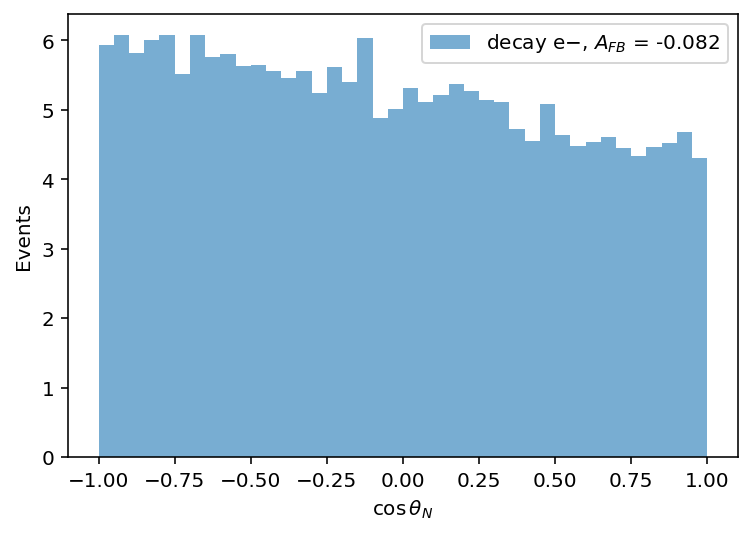}}
    \subfigure[]
    {\includegraphics[width=0.45\textwidth]{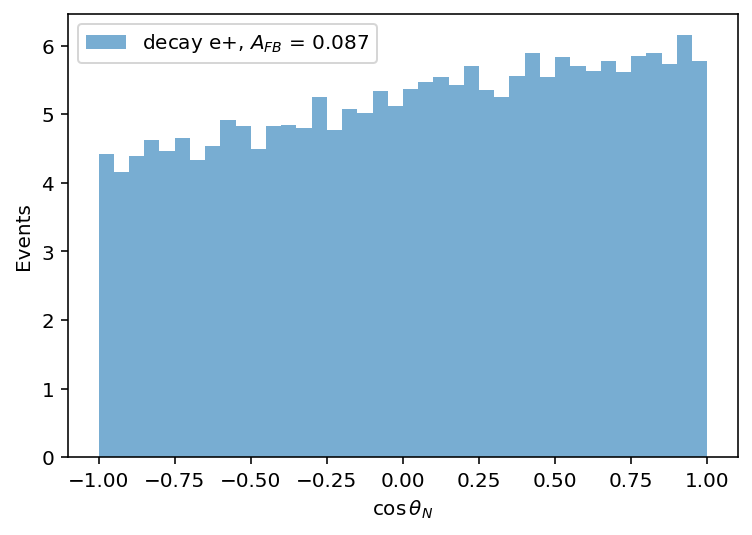}}
    \caption{(a) The angular distributions between the direction of outgoing electrons and the forward hemisphere. (b) The angular distributions between the direction of outgoing positrons and the forward hemisphere.}
    \label{angulardistributionlrsm}
\end{figure}

The chiral interactions of the $W'$ boson (effectively, $W_R$) can be investigated by looking at the Forward-Backward Asymmetry $A_{\rm FB}$ of the final state lepton originating from the heavy neutrino in the latter rest frame. The forward direction is defined as the one of the incoming quark, i.e., the parton  with the highest momentum. We boost the lepton and quarks to the neutrino rest frame and define $\theta_N$ as the angle between the momentum of the final state lepton and the forward direction. Then, by looking at the sign of $\cos \theta_N$, we can extract the $A_{\rm FB}$ value for all the generated events. We further distinguish between positive and negative leptons. 
Since the heavy neutrino mass is much smaller than the $W'$ mass, the chirality of the heavy neutrino will be right-handed. The results are plotted in Fig.~\ref{angulardistributionlrsm}. Such a distribution appears  to offer clear 
diagnostic power of the $W'$ properties.

\begin{figure}
    \centering
    \subfigure[]
    {\includegraphics[width=0.45\textwidth]{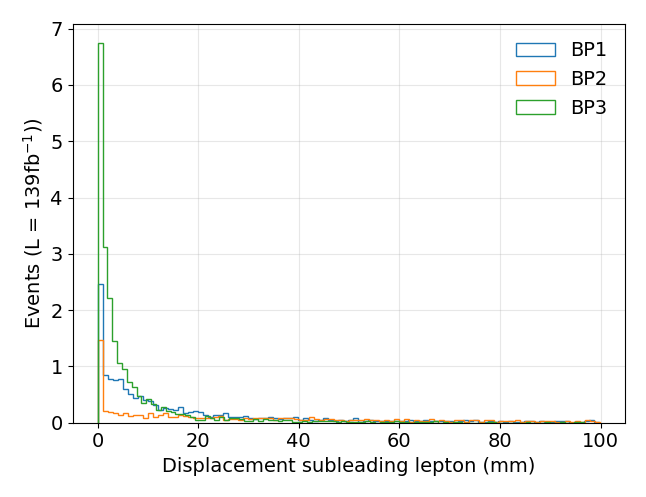}}
    \subfigure[]
    {\includegraphics[width=0.45\textwidth]{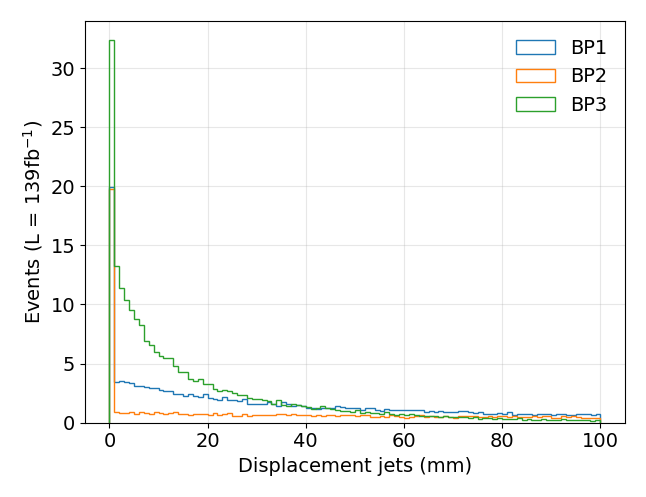}}
    \caption{(a) The total displacement distributions for the sub-leading lepton in each BP. (b) The total displacement distribution for the leading jet in each BP.}
    \label{displacementslrsm}
\end{figure}

To obtain the vertex displacements, a modified version of DELPHES \cite{deFavereau:2013fsa} was used which tracks the decay vertices of the heavy neutrinos and gives us information about the coordinates of the decays vertex which can be used to calculate the total displacement of the objects under consideration. The vertex displacements for the sub-leading lepton and the leading jets are shown for each BP in Fig.~\ref{displacementslrsm}. These distributions show that, for the given BPs, displaced signals can be expected, with the possible exception of BP2 which has a lower $M(N_{e})$ than in the other two BPs. Therefore, the heavy neutrino will be more boosted in BP2 than in the other two BPs which results in the heavy neutrino having a longer lifetime in the lab frame, in turn resulting in a large fraction of the decays happen outside the typical inner detector of the LHC. This can also be seen in the invariant mass distributions. Although, the three BPs have similar cross sections, the number of detected events are lower for BP2 due to its long lifetime, especially in the region where the $W_R$ is produced on-shell,  as the neutrinos are expected to be the most boosted. 

\subsection{Phenomenological Results for $U(1)_{B-L}$}
\begin{figure} [t!]
    \centering
    \includegraphics[width=\linewidth]{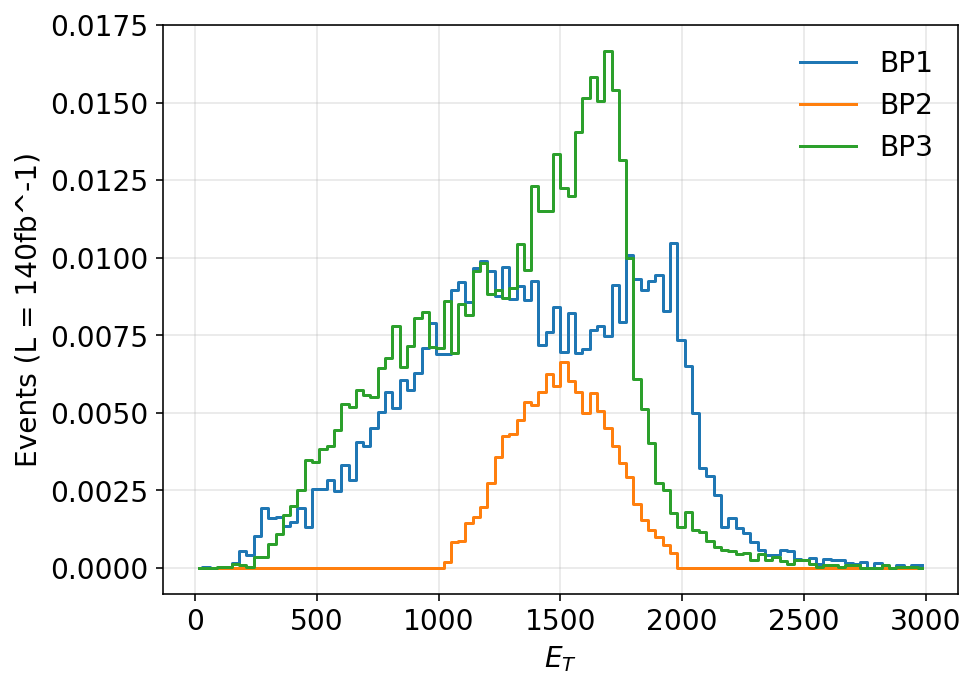}
    \caption{The total transverse energy distribution (in GeV) for the chosen BPs.}
    \label{totalebl}
\end{figure}

Due to the choice of $Z'$ mass in the BPs for the $U(1)_{B-L}$ gauge extension, the leptons and jets in the final state are not as boosted as in the LRSM case. Therefore, we will have more resolvable final state leptons and so we will not specify any SRs. In the process under consideration for this model, 
$p \space p \rightarrow Z \rightarrow N \space  N \rightarrow e^{\pm} \space e^{\pm} \space jj$,
the final state has several jets and, therefore, the process of combining the jets and leptons in useful invariant mass distributions  is a  difficult task. To get around this,  we can plot the total transverse energy, which is displayed in Fig.~\ref{totalebl} for each of the BPs. Since the $Z'$ is mostly produced on-shell we see a peak of the total transverse energy at the mass of the $Z'$ boson and a sharp drop following the peak. Due to the small cross section of this process, we do not expect to see any events at current LHC luminosities but the channel can be of interest at HL-LHC. 

\begin{figure} 
    \centering
    \subfigure[]
    {\includegraphics[width=0.45\textwidth]{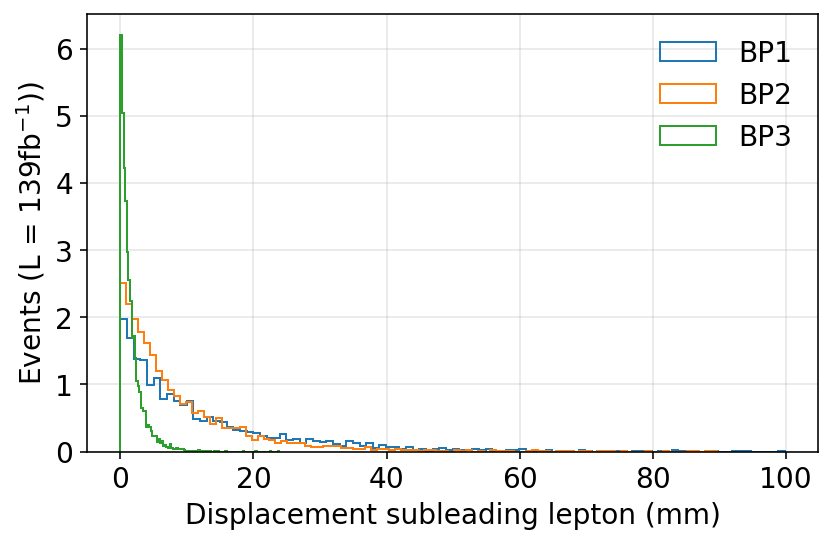}}
    \subfigure[]
    {\includegraphics[width=0.45\textwidth]{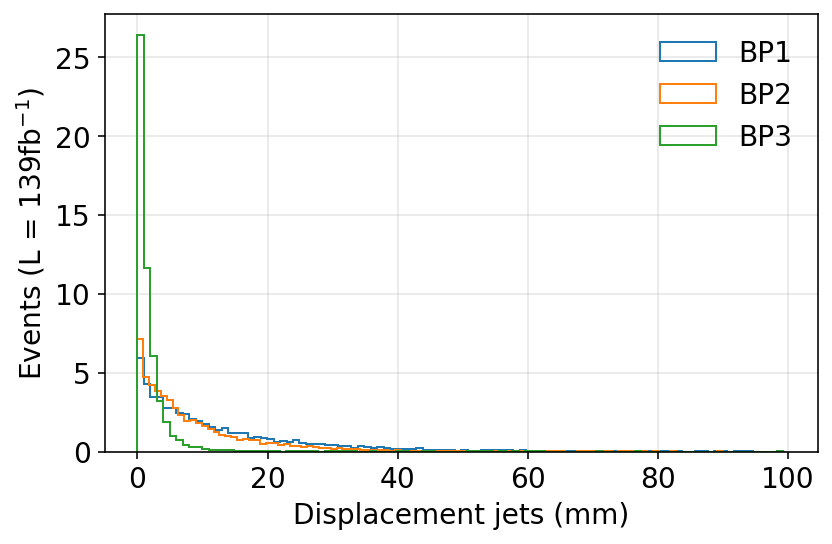}}
    \caption{(a) The total displacement distributions for the sub-leading lepton in each BP. (b) The total displacement distribution for the leading jet in each BP.}
    \label{displacementU1BL}
\end{figure}

\begin{figure} 
    \centering
    \subfigure[]
    {\includegraphics[width=0.45\textwidth]{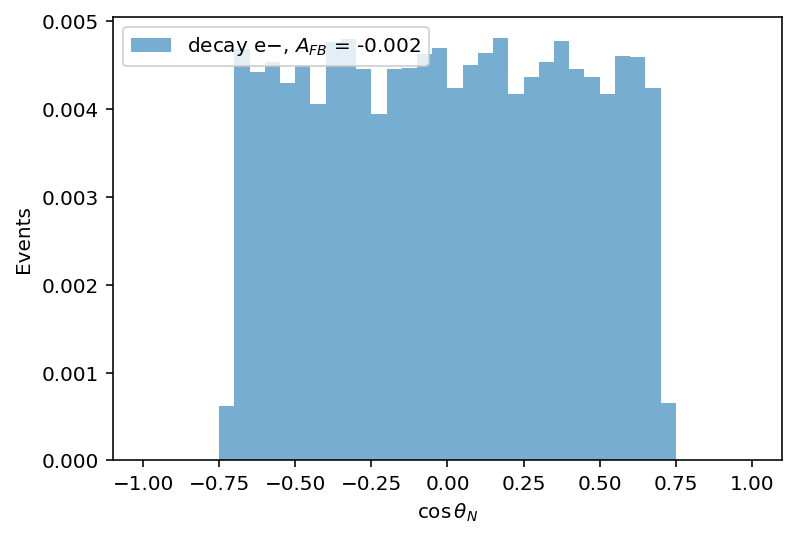}}
    \subfigure[]
    {\includegraphics[width=0.45\textwidth]{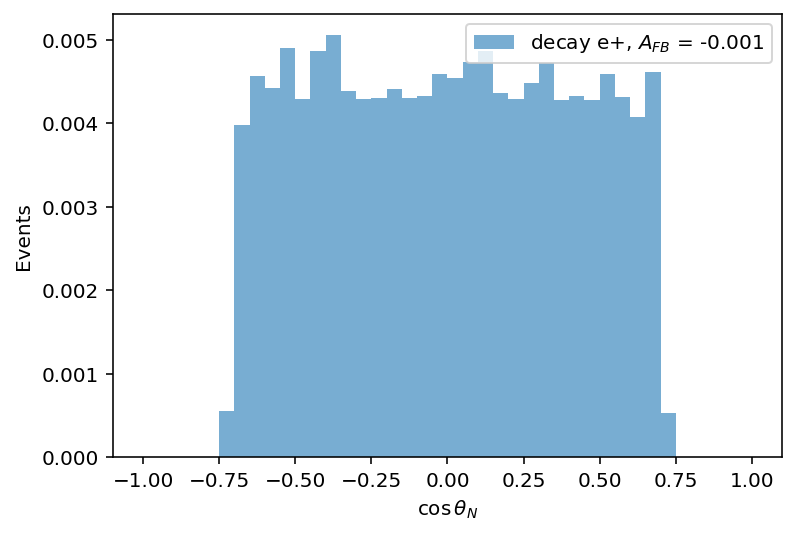}}
    \caption{(a) The angular distributions between the direction of outgoing electrons and the forward hemisphere. (b) The angular distributions between the direction of outgoing positrons and the forward hemisphere.}
    \label{angulardistributionbl}
\end{figure}

For the three BPs previously introduced for the $U(1)_{B-L}$ model, we display the displacement length expected in the detector for the subleading lepton (Fig.~\ref{displacementU1BL}(a)) and leading jet (Fig.~\ref{displacementU1BL}(b)), showing that a degeneracy exist between BP1 and BP2 while BP3 has a distinctive dependence in this variable, which can then be used for diagnostic purposes. 
We finally perform a similar analysis on the $A_{\rm FB}$ of the final state leptons as done in the LRSM. Here, however, the $Z'$ has vector-like couplings to the heavy neutrinos and so the latter are essentially unpolarised, which results in a vanishing asymmetry. These results are plotted in Fig.~\ref{angulardistributionbl}, including both  $Z$ and $Z'$ mediation, which, again, show the scope of the HL-LHC in characterising the nature of the couplings of heavy neutrinos to the predominant $U(1)_{B-L}$ mediator (recall that $Z$ mediation would generate a non-zero $A_{\rm FB}$).

\section{Conclusions}
In summary, we have performed a preliminary analysis, based on parton level results, testing the potential scope of the HL-LHC in accessing signals of
heavy neutrinos of the LRSM and $U(1)_{B-L}$ model when, crucially, the latter are LLPs. Under such a condition, the dominant background is the intrinsic one in both cases, involving a subleading (for large momentum transfer) component due to the $W$ (for the LRSM) and $Z$ (for the $U(1)_{B-L}$ model) bosons of the SM, which therefore do not spoil the ability of charactering both the BSM mediators ($W'$ and $Z'$, respectively) and  heavy neutrino itself, in terms of the masses and couplings involved. 

We are aware that more sophisticated analyses are in order, including parton shower, hadronisation and jet definition, alongside detector effects on these as well as leptons, before claiming any substantive scope by the HL-LHC in extracting our advocated signals. Yet, the fact that all such objects originate from displaced heavy neutrinos in a kinematical region which should be free from other SM backgrounds, gives us confidence in the importance of our results.  For the purpose of aiding such future analyses, we have herein published three BPs in each BSM scenario considered, which fulfil the purpose of offering inclusive rates most probably not accessible at the LHC but  certainly so at the HL-LHC, simultaneously offering various differential distributions (including asymmetries) 
interpretable in terms of the fundamental $W'$, $Z'$ and heavy neutrino properties, while also displaying displaced vertices (for the emerging leptons and jets)  inside the ATLAS and CMS detectors.

\begin{acknowledgments}
SM is supported in part through the NExT Institute and STFC Consolidated Grant ST/X000583/1. 
We all thank Harri Waltari for actual collaboration in the initial stages of this research as  well as for innumerable discussions throughout.
\end{acknowledgments}

\bibliography{paper_refs}

@article{Gianotti:2002xx,
    author = "Gianotti, F. and others",
    title = "{Physics potential and experimental challenges of the LHC luminosity upgrade}",
    eprint = "hep-ph/0204087",
    archivePrefix = "arXiv",
    reportNumber = "CERN-TH-2002-078",
    doi = "10.1140/epjc/s2004-02061-6",
    journal = "Eur. Phys. J. C",
    volume = "39",
    pages = "293--333",
    year = "2005"
}

@article{Super-Kamiokande:1998kpq,
    author = "Fukuda, Y. and others",
    collaboration = "Super-Kamiokande",
    title = "{Evidence for oscillation of atmospheric neutrinos}",
    eprint = "hep-ex/9807003",
    archivePrefix = "arXiv",
    reportNumber = "BU-98-17, ICRR-REPORT-422-98-18, UCI-98-8, KEK-PREPRINT-98-95, LSU-HEPA-5-98, UMD-98-003, SBHEP-98-5, TKU-PAP-98-06, TIT-HPE-98-09",
    doi = "10.1103/PhysRevLett.81.1562",
    journal = "Phys. Rev. Lett.",
    volume = "81",
    pages = "1562--1567",
    year = "1998"
}

@article{LSND:2001aii,
    author = "Aguilar, A. and others",
    collaboration = "LSND",
    title = "{Evidence for neutrino oscillations from the observation of $\bar{\nu}_e$ appearance in a $\bar{\nu}_\mu$
 beam}",
    eprint = "hep-ex/0104049",
    archivePrefix = "arXiv",
    doi = "10.1103/PhysRevD.64.112007",
    journal = "Phys. Rev. D",
    volume = "64",
    pages = "112007",
    year = "2001"
}

@article{K2K:2002icj,
    author = "Ahn, M. H. and others",
    collaboration = "K2K",
    title = "{Indications of neutrino oscillation in a 250 km long baseline experiment}",
    eprint = "hep-ex/0212007",
    archivePrefix = "arXiv",
    doi = "10.1103/PhysRevLett.90.041801",
    journal = "Phys. Rev. Lett.",
    volume = "90",
    pages = "041801",
    year = "2003"
}

@article{T2K:2011ypd,
    author = "Abe, K. and others",
    collaboration = "T2K",
    title = "{Indication of Electron Neutrino Appearance from an Accelerator-produced Off-axis Muon Neutrino Beam}",
    eprint = "1106.2822",
    archivePrefix = "arXiv",
    primaryClass = "hep-ex",
    doi = "10.1103/PhysRevLett.107.041801",
    journal = "Phys. Rev. Lett.",
    volume = "107",
    pages = "041801",
    year = "2011"
}

@article{DayaBay:2013yxg,
    author = "An, F. P. and others",
    collaboration = "Daya Bay",
    title = "{Spectral measurement of electron antineutrino oscillation amplitude and frequency at Daya Bay}",
    eprint = "1310.6732",
    archivePrefix = "arXiv",
    primaryClass = "hep-ex",
    doi = "10.1103/PhysRevLett.112.061801",
    journal = "Phys. Rev. Lett.",
    volume = "112",
    pages = "061801",
    year = "2014"
}

@article{Minkowski:1977sc,
    author = "Minkowski, Peter",
    title = "{$\mu \to e\gamma$ at a Rate of One Out of $10^{9}$ Muon Decays?}",
    reportNumber = "Print-77-0182 (BERN)",
    doi = "10.1016/0370-2693(77)90435-X",
    journal = "Phys. Lett. B",
    volume = "67",
    pages = "421--428",
    year = "1977"
}

@article{Mohapatra:1979ia,
    author = "Mohapatra, Rabindra N. and Senjanovic, Goran",
    title = "{Neutrino Mass and Spontaneous Parity Nonconservation}",
    reportNumber = "MDDP-TR-80-060, MDDP-PP-80-105, CCNY-HEP-79-10",
    doi = "10.1103/PhysRevLett.44.912",
    journal = "Phys. Rev. Lett.",
    volume = "44",
    pages = "912",
    year = "1980"
}

@article{Mohapatra:1980yp,
    author = "Mohapatra, Rabindra N. and Senjanovic, Goran",
    title = "{Neutrino Masses and Mixings in Gauge Models with Spontaneous Parity Violation}",
    reportNumber = "FERMILAB-PUB-80-061-THY, FERMILAB-PUB-80-061-T",
    doi = "10.1103/PhysRevD.23.165",
    journal = "Phys. Rev. D",
    volume = "23",
    pages = "165",
    year = "1981"
}

@article{Gell-Mann:1979vob,
    author = "Gell-Mann, Murray and Ramond, Pierre and Slansky, Richard",
    title = "{Complex Spinors and Unified Theories}",
    eprint = "1306.4669",
    archivePrefix = "arXiv",
    primaryClass = "hep-th",
    reportNumber = "PRINT-80-0576",
    journal = "Conf. Proc. C",
    volume = "790927",
    pages = "315--321",
    year = "1979"
}

@article{Magg:1980ut,
    author = "Magg, M. and Wetterich, C.",
    title = "{Neutrino Mass Problem and Gauge Hierarchy}",
    reportNumber = "CERN-TH-2829",
    doi = "10.1016/0370-2693(80)90825-4",
    journal = "Phys. Lett. B",
    volume = "94",
    pages = "61--64",
    year = "1980"
}

@article{Vergados:1985pq,
    author = "Vergados, J. D.",
    title = "{The Neutrino Mass and Family, Lepton and Baryon Nonconservation in Gauge Theories}",
    reportNumber = "IOANNINA-187",
    doi = "10.1016/0370-1573(86)90088-8",
    journal = "Phys. Rept.",
    volume = "133",
    pages = "1",
    year = "1986"
}

@article{Pati:1974yy,
    author = "Pati, Jogesh C. and Salam, Abdus",
    title = "{Lepton Number as the Fourth Color}",
    reportNumber = "IC-74-7",
    doi = "10.1103/PhysRevD.10.275",
    journal = "Phys. Rev. D",
    volume = "10",
    pages = "275--289",
    year = "1974",
    note = "[Erratum: Phys.Rev.D 11, 703--703 (1975)]"
}

@article{Mohapatra:1974hk,
    author = "Mohapatra, Rabindra N. and Pati, Jogesh C.",
    title = "{Left-Right Gauge Symmetry and an Isoconjugate Model of CP Violation}",
    reportNumber = "MDDP-TR-74-085",
    doi = "10.1103/PhysRevD.11.566",
    journal = "Phys. Rev. D",
    volume = "11",
    pages = "566--571",
    year = "1975"
}

@article{Senjanovic:1975rk,
    author = "Senjanovic, G. and Mohapatra, Rabindra N.",
    title = "{Exact Left-Right Symmetry and Spontaneous Violation of Parity}",
    reportNumber = "CCNY-HEP-75-5",
    doi = "10.1103/PhysRevD.12.1502",
    journal = "Phys. Rev. D",
    volume = "12",
    pages = "1502",
    year = "1975"
}

@article{Cottin:2018kmq,
    author = "Cottin, Giovanna and Helo, Juan Carlos and Hirsch, Martin",
    title = "{Searches for light sterile neutrinos with multitrack displaced vertices}",
    eprint = "1801.02734",
    archivePrefix = "arXiv",
    primaryClass = "hep-ph",
    doi = "10.1103/PhysRevD.97.055025",
    journal = "Phys. Rev. D",
    volume = "97",
    number = "5",
    pages = "055025",
    year = "2018"
}

@article{Keung:1983uu,
    author = "Keung, Wai-Yee and Senjanovic, Goran",
    title = "{Majorana Neutrinos and the Production of the Right-handed Charged Gauge Boson}",
    reportNumber = "BNL-32872",
    doi = "10.1103/PhysRevLett.50.1427",
    journal = "Phys. Rev. Lett.",
    volume = "50",
    pages = "1427",
    year = "1983"
}

@article{Nemevsek:2018bbt,
    author = "Nemev{\v{s}}ek, Miha and Nesti, Fabrizio and Popara, Goran",
    title = "{Keung-Senjanovi{\'c} process at the LHC: From lepton number violation to displaced vertices to invisible decays}",
    eprint = "1801.05813",
    archivePrefix = "arXiv",
    primaryClass = "hep-ph",
    doi = "10.1103/PhysRevD.97.115018",
    journal = "Phys. Rev. D",
    volume = "97",
    number = "11",
    pages = "115018",
    year = "2018"
}

@article{Urquia-Calderon:2023dkf,
    author = "Urqu{\'\i}a-Calder{\'o}n, Kevin A.",
    title = "{Long-lived heavy neutral leptons at lepton colliders as a probe of left-right-symmetric models}",
    eprint = "2310.17406",
    archivePrefix = "arXiv",
    primaryClass = "hep-ph",
    doi = "10.1103/PhysRevD.109.055002",
    journal = "Phys. Rev. D",
    volume = "109",
    number = "5",
    pages = "055002",
    year = "2024"
}

@article{Basso:2008iv,
    author = "Basso, Lorenzo and Belyaev, Alexander and Moretti, Stefano and Shepherd-Themistocleous, Claire H.",
    title = "{Phenomenology of the minimal $B-L$ extension of the Standard model: Z' and neutrinos}",
    eprint = "0812.4313",
    archivePrefix = "arXiv",
    primaryClass = "hep-ph",
    reportNumber = "SHEP-08-13",
    doi = "10.1103/PhysRevD.80.055030",
    journal = "Phys. Rev. D",
    volume = "80",
    pages = "055030",
    year = "2009"
}

@article{Bandyopadhyay:2022mej,
    author = "Bandyopadhyay, Priyotosh and Chun, Eung Jin and Sen, Chandrima",
    title = "{Boosted displaced decay of right-handed neutrinos at CMS, ATLAS and MATHUSLA}",
    eprint = "2205.12511",
    archivePrefix = "arXiv",
    primaryClass = "hep-ph",
    reportNumber = "IITH-PH-0002/22, KIAS-P22017",
    doi = "10.1007/JHEP02(2023)103",
    journal = "JHEP",
    volume = "02",
    pages = "103",
    year = "2023"
}

@article{Wetterich:1981bx,
    author = "Wetterich, C.",
    title = "{Neutrino Masses and the Scale of $B-L$ Violation}",
    reportNumber = "FREIBURG-THEP-81-2",
    doi = "10.1016/0550-3213(81)90279-0",
    journal = "Nucl. Phys. B",
    volume = "187",
    pages = "343--375",
    year = "1981"
}

@article{Mohapatra:1980qe,
    author = "Mohapatra, Rabindra N. and Marshak, R. E.",
    title = "{Local $B-L$ Symmetry of Electroweak Interactions, Majorana Neutrinos and Neutron Oscillations}",
    reportNumber = "VPI-HEP-80/1",
    doi = "10.1103/PhysRevLett.44.1316",
    journal = "Phys. Rev. Lett.",
    volume = "44",
    pages = "1316--1319",
    year = "1980",
    note = "[Erratum: Phys.Rev.Lett. 44, 1643 (1980)]"
}

@article{Buchmuller:1991ce,
    author = "Buchmuller, W. and Greub, C. and Minkowski, P.",
    title = "{Neutrino masses, neutral vector bosons and the scale of $B-L$ breaking}",
    reportNumber = "DESY-91-053",
    doi = "10.1016/0370-2693(91)90952-M",
    journal = "Phys. Lett. B",
    volume = "267",
    pages = "395--399",
    year = "1991"
}

@article{Khalil:2006yi,
    author = "Khalil, Shaaban",
    title = "{Low scale $B - L$ extension of the Standard Model at the LHC}",
    eprint = "hep-ph/0611205",
    archivePrefix = "arXiv",
    doi = "10.1088/0954-3899/35/5/055001",
    journal = "J. Phys. G",
    volume = "35",
    pages = "055001",
    year = "2008"
}

@article{ATLAS:2025uah,
    author = "Aad, Georges and others",
    collaboration = "ATLAS",
    title = "{Search for heavy neutral leptons in decays of W bosons using leptonic and semi-leptonic displaced vertices in $ \sqrt{s} $ = 13 TeV pp collisions with the ATLAS detector}",
    eprint = "2503.16213",
    archivePrefix = "arXiv",
    primaryClass = "hep-ex",
    reportNumber = "CERN-EP-2025-052",
    doi = "10.1007/JHEP07(2025)196",
    journal = "JHEP",
    volume = "07",
    pages = "196",
    year = "2025"
}

@article{CMS:2024hik,
    author = "Hayrapetyan, Aram and others",
    collaboration = "CMS",
    title = "{Search for long-lived heavy neutral leptons in proton-proton collision events with a lepton-jet pair associated with a secondary vertex at $ \sqrt{s} $ = 13 TeV}",
    eprint = "2407.10717",
    archivePrefix = "arXiv",
    primaryClass = "hep-ex",
    reportNumber = "CMS-EXO-21-011, CERN-EP-2024-161",
    doi = "10.1007/JHEP02(2025)036",
    journal = "JHEP",
    volume = "02",
    pages = "036",
    year = "2025"
}

@phdthesis{Liu:2020nqi,
    author = "Liu, Wei",
    title = "{Exploring B-L gauge models at the LHC and beyond}",
    school = "University Coll. London",
    year = "2020"
}

@article{CMS:2018agk,
    author = "Sirunyan, Albert M and others",
    collaboration = "CMS",
    title = "{Search for a heavy right-handed W boson and a heavy neutrino in events with two same-flavor leptons and two jets at $\sqrt{s}=$ 13 TeV}",
    eprint = "1803.11116",
    archivePrefix = "arXiv",
    primaryClass = "hep-ex",
    reportNumber = "CMS-EXO-17-011, CERN-EP-2018-028",
    doi = "10.1007/JHEP05(2018)148",
    journal = "JHEP",
    volume = "05",
    pages = "148",
    year = "2018"
}

@article{CMS:2021dzb,
    author = "Tumasyan, Armen and others",
    collaboration = "CMS",
    title = "{Search for a right-handed W boson and a heavy neutrino in proton-proton collisions at $ \sqrt{s} $ = 13 TeV}",
    eprint = "2112.03949",
    archivePrefix = "arXiv",
    primaryClass = "hep-ex",
    reportNumber = "CMS-EXO-20-002, CERN-EP-2021-228",
    doi = "10.1007/JHEP04(2022)047",
    journal = "JHEP",
    volume = "04",
    pages = "047",
    year = "2022"
}

@article{ATLAS:2015gtp,
    author = "Aad, Georges and others",
    collaboration = "ATLAS",
    title = "{Search for heavy Majorana neutrinos with the ATLAS detector in pp collisions at $ \sqrt{s}=8 $ TeV}",
    eprint = "1506.06020",
    archivePrefix = "arXiv",
    primaryClass = "hep-ex",
    reportNumber = "CERN-PH-EP-2015-070",
    doi = "10.1007/JHEP07(2015)162",
    journal = "JHEP",
    volume = "07",
    pages = "162",
    year = "2015"
}

@article{ATLAS:2018dcj,
    author = "Aaboud, Morad and others",
    collaboration = "ATLAS",
    title = "{Search for heavy Majorana or Dirac neutrinos and right-handed $W$ gauge bosons in final states with two charged leptons and two jets at $ \sqrt{s}=13 $ TeV with the ATLAS detector}",
    eprint = "1809.11105",
    archivePrefix = "arXiv",
    primaryClass = "hep-ex",
    reportNumber = "CERN-EP-2018-199",
    doi = "10.1007/JHEP01(2019)016",
    journal = "JHEP",
    volume = "01",
    pages = "016",
    year = "2019"
}

@article{ATLAS:2019isd,
    author = "Aaboud, Morad and others",
    collaboration = "ATLAS",
    title = "{Search for a right-handed gauge boson decaying into a high-momentum heavy neutrino and a charged lepton in $pp$ collisions with the ATLAS detector at $\sqrt{s}=13$ TeV}",
    eprint = "1904.12679",
    archivePrefix = "arXiv",
    primaryClass = "hep-ex",
    reportNumber = "CERN-EP-2019-052",
    doi = "10.1016/j.physletb.2019.134942",
    journal = "Phys. Lett. B",
    volume = "798",
    pages = "134942",
    year = "2019"
}

@article{CMS:2019gwf,
    author = "Sirunyan, Albert M and others",
    collaboration = "CMS",
    title = "{Search for high mass dijet resonances with a new background prediction method in proton-proton collisions at $\sqrt{s} =$ 13 TeV}",
    eprint = "1911.03947",
    archivePrefix = "arXiv",
    primaryClass = "hep-ex",
    reportNumber = "CMS-EXO-19-012, CERN-EP-2019-222",
    doi = "10.1007/JHEP05(2020)033",
    journal = "JHEP",
    volume = "05",
    pages = "033",
    year = "2020"
}

@article{ATLAS:2023cjo,
    author = "Aad, Georges and others",
    collaboration = "ATLAS",
    title = "{Search for heavy Majorana or Dirac neutrinos and right-handed W gauge bosons in final states with charged leptons and jets in pp collisions at $\sqrt{s}=13$~TeV with the ATLAS detector}",
    eprint = "2304.09553",
    archivePrefix = "arXiv",
    primaryClass = "hep-ex",
    reportNumber = "CERN-EP-2023-034",
    doi = "10.1140/epjc/s10052-023-12021-9",
    journal = "Eur. Phys. J. C",
    volume = "83",
    number = "12",
    pages = "1164",
    year = "2023"
}

@article{Porod:2003um,
    author = "Porod, Werner",
    title = "{SPheno, a program for calculating supersymmetric spectra, SUSY particle decays and SUSY particle production at $e^+ e^-$ colliders}",
    eprint = "hep-ph/0301101",
    archivePrefix = "arXiv",
    reportNumber = "ZU-TH-01-03",
    doi = "10.1016/S0010-4655(03)00222-4",
    journal = "Comput. Phys. Commun.",
    volume = "153",
    pages = "275--315",
    year = "2003"
}

@article{Porod:2011nf,
    author = "Porod, W. and Staub, F.",
    title = "{SPheno 3.1: Extensions including flavour, CP-phases and models beyond the MSSM}",
    eprint = "1104.1573",
    archivePrefix = "arXiv",
    primaryClass = "hep-ph",
    doi = "10.1016/j.cpc.2012.05.021",
    journal = "Comput. Phys. Commun.",
    volume = "183",
    pages = "2458--2469",
    year = "2012"
}

@article{Staub:2013tta,
    author = "Staub, Florian",
    title = "{SARAH 4: A tool for (not only SUSY) model builders}",
    eprint = "1309.7223",
    archivePrefix = "arXiv",
    primaryClass = "hep-ph",
    reportNumber = "BONN-TH-2013-17",
    doi = "10.1016/j.cpc.2014.02.018",
    journal = "Comput. Phys. Commun.",
    volume = "185",
    pages = "1773--1790",
    year = "2014"
}

@article{Staub:2008uz,
    author = "Staub, F.",
    title = "{SARAH}",
    eprint = "0806.0538",
    archivePrefix = "arXiv",
    primaryClass = "hep-ph",
    month = "arXiv:0806.0538",
    year = "",
    journal = ""
}

@article{Staub:2009bi,
    author = "Staub, Florian",
    title = "{From Superpotential to Model Files for FeynArts and CalcHep/CompHep}",
    eprint = "0909.2863",
    archivePrefix = "arXiv",
    primaryClass = "hep-ph",
    doi = "10.1016/j.cpc.2010.01.011",
    journal = "Comput. Phys. Commun.",
    volume = "181",
    pages = "1077--1086",
    year = "2010"
}

@article{Staub:2010jh,
    author = "Staub, Florian",
    title = "{Automatic Calculation of supersymmetric Renormalization Group Equations and Self Energies}",
    eprint = "1002.0840",
    archivePrefix = "arXiv",
    primaryClass = "hep-ph",
    doi = "10.1016/j.cpc.2010.11.030",
    journal = "Comput. Phys. Commun.",
    volume = "182",
    pages = "808--833",
    year = "2011"
}

@article{Staub:2012pb,
    author = "Staub, Florian",
    title = "{SARAH 3.2: Dirac Gauginos, UFO output, and more}",
    eprint = "1207.0906",
    archivePrefix = "arXiv",
    primaryClass = "hep-ph",
    reportNumber = "BONN-TH-2012-17",
    doi = "10.1016/j.cpc.2013.02.019",
    journal = "Comput. Phys. Commun.",
    volume = "184",
    pages = "1792--1809",
    year = "2013"
}

@article{Bonilla:2016fqd,
    author = "Bonilla, Cesar and Krauss, Manuel E. and Opferkuch, Toby and Porod, Werner",
    title = "{Perspectives for Detecting Lepton Flavour Violation in Left-Right Symmetric Models}",
    eprint = "1611.07025",
    archivePrefix = "arXiv",
    primaryClass = "hep-ph",
    reportNumber = "BONN-TH-2016-08, IFIC-16-XX",
    doi = "10.1007/JHEP03(2017)027",
    journal = "JHEP",
    volume = "03",
    pages = "027",
    year = "2017"
}

@article{Alwall:2014hca,
    author = "Alwall, J. and Frederix, R. and Frixione, S. and Hirschi, V. and Maltoni, F. and Mattelaer, O. and Shao, H. -S. and Stelzer, T. and Torrielli, P. and Zaro, M.",
    title = "{The automated computation of tree-level and next-to-leading order differential cross sections, and their matching to parton shower simulations}",
    eprint = "1405.0301",
    archivePrefix = "arXiv",
    primaryClass = "hep-ph",
    reportNumber = "CERN-PH-TH-2014-064, CP3-14-18, LPN14-066, MCNET-14-09, ZU-TH-14-14",
    doi = "10.1007/JHEP07(2014)079",
    journal = "JHEP",
    volume = "07",
    pages = "079",
    year = "2014"
}

@article{Ito:2017dpm,
    author = "Ito, Hayato and Jinnouchi, Osamu and Moroi, Takeo and Nagata, Natsumi and Otono, Hidetoshi",
    title = "{Extending the LHC Reach for New Physics with Sub-Millimeter Displaced Vertices}",
    eprint = "1702.08613",
    archivePrefix = "arXiv",
    primaryClass = "hep-ph",
    reportNumber = "UT-17-07, KYUSHU-RCAPP-2017-02",
    doi = "10.1016/j.physletb.2017.06.003",
    journal = "Phys. Lett. B",
    volume = "771",
    pages = "568--575",
    year = "2017"
}

@article{CMS:2013vyz,
    author = "Chatrchyan, Serguei and others",
    collaboration = "CMS",
    title = "{The Performance of the CMS Muon Detector in Proton-Proton Collisions at $\sqrt{s}$ = 7 TeV at the LHC}",
    eprint = "1306.6905",
    archivePrefix = "arXiv",
    primaryClass = "physics.ins-det",
    reportNumber = "CMS-MUO-11-001, CERN-PH-EP-2013-072",
    doi = "10.1088/1748-0221/8/11/P11002",
    journal = "JINST",
    volume = "8",
    pages = "P11002",
    year = "2013"
}

@article{ATLAS:2019erb,
    author = "Aad, Georges and others",
    collaboration = "ATLAS",
    title = "{Search for high-mass dilepton resonances using 139 fb$^{-1}$ of $pp$ collision data collected at $\sqrt{s}=$13 TeV with the ATLAS detector}",
    eprint = "1903.06248",
    archivePrefix = "arXiv",
    primaryClass = "hep-ex",
    reportNumber = "CERN-EP-2019-030",
    doi = "10.1016/j.physletb.2019.07.016",
    journal = "Phys. Lett. B",
    volume = "796",
    pages = "68--87",
    year = "2019"
}

@article{ATLAS:2020fry,
    author = "Aad, Georges and others",
    collaboration = "ATLAS",
    title = "{Search for heavy diboson resonances in semileptonic final states in pp collisions at $\sqrt{s}=13$ TeV with the ATLAS detector}",
    eprint = "2004.14636",
    archivePrefix = "arXiv",
    primaryClass = "hep-ex",
    reportNumber = "CERN-EP-2020-049",
    doi = "10.1140/epjc/s10052-020-08554-y",
    journal = "Eur. Phys. J. C",
    volume = "80",
    number = "12",
    pages = "1165",
    year = "2020"
}

@article{Accomando:2013sfa,
    author = "Accomando, Elena and Becciolini, Diego and Belyaev, Alexander and Moretti, Stefano and Shepherd-Themistocleous, Claire",
    title = "{$Z'$ at the LHC: Interference and Finite Width Effects in Drell-Yan}",
    eprint = "1304.6700",
    archivePrefix = "arXiv",
    primaryClass = "hep-ph",
    reportNumber = "CP3-ORIGINS-2013-013, DIAS-2013-13, SHEP-13-02",
    doi = "10.1007/JHEP10(2013)153",
    journal = "JHEP",
    volume = "10",
    pages = "153",
    year = "2013"
}

@article{Accomando:2016sge,
    author = "Accomando, Elena and Corian\'o, Claudio and Delle Rose, Luigi and Fiaschi, Juri and Marzo, Carlo and Moretti, Stefano",
    title = "{$Z^{'}$, Higgses and heavy neutrinos in $U(1)^{'}$ models: from the LHC to the GUT scale}",
    eprint = "1605.02910",
    archivePrefix = "arXiv",
    primaryClass = "hep-ph",
    doi = "10.1007/JHEP07(2016)086",
    journal = "JHEP",
    volume = "07",
    pages = "086",
    year = "2016"
}

@article{Accomando:2017qcs,
    author = "Accomando, Elena and Delle Rose, Luigi and Moretti, Stefano and Olaiya, Emmanuel and Shepherd-Themistocleous, Claire H.",
    title = "{Extra Higgs boson and $Z^{'}$ as portals to signatures of heavy neutrinos at the LHC}",
    eprint = "1708.03650",
    archivePrefix = "arXiv",
    primaryClass = "hep-ph",
    doi = "10.1007/JHEP02(2018)109",
    journal = "JHEP",
    volume = "02",
    pages = "109",
    year = "2018"
}

@article{deFavereau:2013fsa,
    author = "de Favereau, J. and Delaere, C. and Demin, P. and Giammanco, A. and Lema{\^\i}tre, V. and Mertens, A. and Selvaggi, M.",
    collaboration = "DELPHES 3",
    title = "{DELPHES 3, A modular framework for fast simulation of a generic collider experiment}",
    eprint = "1307.6346",
    archivePrefix = "arXiv",
    primaryClass = "hep-ex",
    doi = "10.1007/JHEP02(2014)057",
    journal = "JHEP",
    volume = "02",
    pages = "057",
    year = "2014"
}

\end{document}